\documentclass[a4paper,11pt]{article}
\usepackage[english]{babel}
\usepackage[utf8x]{inputenc}
\usepackage{amsmath}
\usepackage{amsthm}
\usepackage{amsfonts}
\usepackage{graphicx}
\usepackage{algorithm}
\usepackage[noend]{algpseudocode}
\usepackage[margin=1.1in]{geometry}
\usepackage{microtype}
\usepackage{bbm}
\newtheorem{lemma}{Lemma}
\newtheorem{theorem}{Theorem}
\newtheorem*{theorem*}{Theorem}
\newcommand{\mysize}{\small}
\newcommand{\prob}{\operatorname{Pr}}
\newcommand{\var}{\operatorname{Var}}
\newcommand{\kerr}{k}
\title{\vspace*{-35pt}\textbf{Simple set cardinality estimation\\ through random sampling}\vspace*{10pt}}

\author{
{\mysize Marco Bressan}\\
{\mysize Sapienza Univ.\ Roma}\\
{\mysize bressan@di.uniroma1.it}
\and
{\mysize Enoch Peserico}\\
{\mysize Univ. Padova}\\
{\mysize enoch@dei.unipd.it}
\and
{\mysize Luca Pretto}\\
{\mysize Univ. Padova}\\
{\mysize pretto@dei.unipd.it}
}

\date{}

\begin{document}
\newcommand{\E}{\mathbb{E}}
\newcommand{\rs}{\textsc{sample}}
\newcommand{\cardest}{\textsc{cardapprox}}

\maketitle
\begin{abstract}
\noindent We present a simple algorithm that estimates the cardinality $n$ of a set $V$ when allowed to sample elements of $V$  uniformly and independently at random.
Our algorithm with probability $(1-\delta)$ returns a $(1\pm\epsilon)-$approximation of $n$ drawing $O\big(\sqrt{n} \cdot \epsilon^{-1}\sqrt{\log(\delta^{-1})}\big)$ samples (for $\epsilon^{-1}\sqrt{\log(\delta^{-1})} = O(\sqrt{n})$).
\end{abstract}
\vspace*{25pt}
Tasks like graph-size estimation (see e.g.~\cite{Hardiman&2015} and \cite{Katzir&2014}) have recently revived interest in the problem of estimating the cardinality of a set via random sampling. This short note presents a simple algorithm that estimates with a given precision the cardinality of a set, with a given probability of error. Although the basic estimator we use has been known for a long time (see~\cite{Schnabel1938} and~\cite{Goodman1953}), we leverage a more recent martingale technique to obtain guarantees on the number of samples yielding the desired precision and error probability. Our algorithm and bounds can then be easily used ``black box'' in the design and analysis of other algorithms.

\section{Estimating set cardinality via random sampling}
The algorithm below estimates
the cardinality $n$ of a set $V$ 
through
repeated invocations of a primitive \rs($V$) that, on each invocation, returns an element of $V$ chosen uniformly and independently at random. We formally state and prove the bounds on the probability that the estimate $\hat{n}$ of $n$ is not accurate within a factor $(1\pm\epsilon)$, or that \rs($V$) is invoked ``too many'' times.
We assume  $\epsilon < 1$; otherwise the trivial estimate $\hat{n}=0$ suffices.

\renewcommand{\thealgorithm}{}
\begin{algorithm}[h!]
\caption{\cardest($V, \epsilon, \delta$)}

\begin{algorithmic}[1]
\State $S \leftarrow \emptyset$ \Comment{subset of $V$ seen so far}
\State $w\leftarrow 0$ \Comment{samples taken, each weighted by $|S|$ when taken}
\State $r\leftarrow 0$ \Comment{number of repeats}
\State $\kerr \leftarrow \big\lceil \frac{2 + 4.4\epsilon}{\epsilon^2}\ln\!\big(\frac{3}{\delta}\big) \big\rceil$ \Comment{stopping threshold on the number of repeats}
\Repeat
\State $w\leftarrow w+|S|$
\State $e \leftarrow \rs(V)$
\State $r \leftarrow r + |S \cap e|$
\State $S \leftarrow S \cup e$
\Until $r \geq \kerr$
\State \textbf{return} $\frac{w}{r}$
\end{algorithmic}
\end{algorithm}

\begin{theorem}
\label{thm:cardinality_estimator}
\cardest$(V,\epsilon,\delta)$ with probability greater than $(1-\delta)$ returns an estimate $\hat{n}$ such that 
$(1-\epsilon)n \le \hat{n}\leq(1+\epsilon)n$ and invokes \rs$(V)$ at most $\min(n,2\lceil\!\sqrt{\kerr n}\big\rceil)+k$ times.
\end{theorem}
\begin{proof}
We first show that $\prob[|\hat{n}-n| > \epsilon n] < 2\delta/3$.
We use a martingale tail inequality originally from~\cite{Freedman1975} and stated (and proved) in the following form as Theorem 2.2 of~\cite{Alon&2010}, p.~8:
\begin{theorem}[\cite{Alon&2010}, Theorem 2.2]
\label{thm:alon}
Let $(Z_0,Z_1,\ldots)$ be a martingale with respect to the filter $(\mathcal{F}_i)$. Suppose that $Z_{i+1}-Z_i \le M$ for all $i$, and write $V_t = \sum_{i=1}^t \var(Z_i|\mathcal{F}_{i-1})$. Then for any $z,v>0$ we have
\[
\prob\!\big[Z_t \ge Z_0 + z, V_t \le v \text{ for some } t\big] \le \exp\!{\Big(\!-\frac{z^2}{2(v+Mz)}\Big)}
\]
\end{theorem}
\noindent Let us plug into the formula of Theorem~\ref{thm:alon} the appropriate quantities from \cardest:
\begin{itemize}\itemsep0pt
\item For all $i \ge 1$ let $X_i \in V$ be the $i$-{th} sample, i.e.\ the value of $e$ set by the $i$-th execution of line 7.
\item For all $i \ge 0$ let $\mathcal{F}_i$ be the event space generated by $X_1, \dots, X_i$ -- so that for any random variable $Y$, with $\E[Y|\mathcal{F}_i]$ we mean $\E[Y|X_1,\dots,X_i]$ and with $\var[Y|\mathcal{F}_i]$ we mean $\var[Y|X_1,\dots,X_i]$.
\item For all $i \ge 1$ let $\chi_i = \mathbbm{1}[X_i \in \bigcup_{j=1}^{i-1}\{X_j\}]$ be the indicator variable of the event that the $i$-th sample is a repeat, i.e.\ that it coincides with some previous sample.
\item For all $i \ge 1$ let $P_i = \E[\chi_i|\mathcal{F}_{i-1}] = \frac{1}{n} |\bigcup_{j=1}^{i-1}\{X_j\} |$ be the probability that the $i$-th sample is a repeat, as a function of all previous samples.
\item Let $Z_0 = 0$, and for all $i \ge 1$ let $Z_i=\sum_{j=1}^i (\chi_j - P_{j})$.
It is easy to see that $(Z_i)_{i \ge 0}$ is a martingale with respect to the filter $(\mathcal{F}_i)_{i\ge 0}$, since $Z_i$ is obtained by adding to $Z_{i-1}$ the indicator variable $\chi_i$ and subtracting $P_i$ (i.e.\ its expectation in $\mathcal{F}_{i-1}$).
More formally, $\E[Z_i|\mathcal{F}_{i-1}]=\E[Z_{i-1} + \chi_i- P_i|\mathcal{F}_{i-1}] = Z_{i-1} + (\E[\chi_i|\mathcal{F}_{i-1}]-P_i) = Z_{i-1}$.
\item Let $M=1$, noting that $|Z_{i+1}-Z_i| = |\chi_{i+1} - P_{i+1}| \le 1$ for all $i$.
\end{itemize}
\vspace{5pt}
Finally, note that $\var[Z_j|\mathcal{F}_{j-1}] = \var[\chi_j|\mathcal{F}_{j-1}]$, since $Z_j = Z_{j-1} + \chi_j - P_j$, and $Z_{j-1}$ and $P_j$ are  functions of $X_1,\dots,X_{j-1}$.
Since $\var[\chi_j|\mathcal{F}_{j-1}] = P_j(1-P_j) \le P_j$, we have $V_i = \sum_{j=1}^i \var[Z_j|\mathcal{F}_{j-1}] \le \sum_{j=1}^i P_j$.
Theorem~\ref{thm:alon} then yields the following:
\begin{lemma}
\label{cor:martbound}
For all $z,v > 0$ we have
\begin{align}
\prob\!\Big[Z_i \ge z, \sum_{j=1}^i P_j \le v \text{ for some } i\Big] &\le 
\exp{\!\Big(\!-\frac{z^2}{2(v+z)}\Big)}
\end{align}
\end{lemma}
\noindent
Let us now focus on \cardest.
Note that $\sum_{j=1}^i \chi_j$ and $\sum_{j=1}^i P_j$ are respectively the values of $r$ and of $\frac{w}{n}$ just after the cycle has been executed for the $i$-th time.
Therefore $Z_i$ is the value of $r-\frac{w}{n}$ just after the cycle has been executed for the $i$-th time.
Suppose now that \cardest\ returns $\frac{w}{r} \le n(1-\epsilon)$.
This event implies $r - \frac{w}{n} \ge \epsilon r$ and $\frac{w}{n} \le (1-\epsilon)r$; which in turn implies $Z_i \ge \epsilon\kerr$ and $\sum_{j=1}^i P_j \le (1-\epsilon)\kerr$, since $r - \frac{w}{n}=Z_i$ and $\frac{w}{n}=\sum_{j=1}^i P_j$, and $r=\kerr$ when \cardest\ returns.
Invoking Lemma~\ref{cor:martbound} with $z=\epsilon\kerr$ and $v=(1-\epsilon)\kerr$ yields:
\begin{align}
\prob\!\Big[\frac{w}{r} \le n(1-\epsilon)\Big]  &\le \exp{\!\Big(\!-\frac{\epsilon^2 \kerr^2}{2(\epsilon \kerr + (1-\epsilon)\kerr)}\Big)}
= \exp{\!\Big(\!-\frac{\epsilon^2 \kerr}{2}\Big)}
\end{align}
which is smaller than $\delta/3$ since $\kerr > \frac{2}{\epsilon^2}\ln{\frac{3}{\delta}}$.

Consider now the event that \cardest\ returns $\frac{w}{r} \ge n(1+\epsilon)$, implying $\frac{w}{n} \ge r (1+\epsilon)$.
This means at return time $Z_i \le -\epsilon\kerr$ or, equivalently, $-Z_i \ge \epsilon\kerr$.
Clearly, $(-Z_i)_{i \ge 0}$ is a martingale too with respect to the filter $(\mathcal{F}_i)_{i\ge 0}$.
Let then $i_0 = \min\{j : -Z_{j} \ge \epsilon\kerr\}$.
Since $|Z_j - Z_{j-1}| \le 1$, it must be $-Z_{i_0} < \epsilon\kerr+1$.
Furthermore, $\sum_{j=1}^{i_0} \chi_j \le \kerr$, or \cardest\ would have stopped at time $i' < i_0 \le i$.
It follows that $\sum_{j=1}^{i_0} P_j = -Z_{i_0} + \sum_{j=0}^{i_0} \chi_j \le \epsilon\kerr + 1 + \kerr = (1+\epsilon)\kerr+1$.
Invoking again Lemma~\ref{cor:martbound} with $z=\epsilon\kerr$ and $v=(1+\epsilon)\kerr+1$, we obtain:
\begin{align}
\prob\!\Big[\frac{w}{r} \ge n(1+\epsilon)\Big] 
&\le \exp{\!\Big(\!-\frac{\epsilon^2 \kerr^2}{2((1+2\epsilon)\kerr+1)}\Big)}
\end{align}
Note that $\frac{1}{\kerr} < \frac{\epsilon^2}{2+4.4\epsilon} < 0.2\epsilon$ since $\epsilon < 1$; so $2((1+2\epsilon)+\frac{1}{\kerr}) < 2+4.4\epsilon$, and since $\kerr \ge \frac{2+4.4\epsilon}{\epsilon^2}\ln{\!\frac{3}{\delta}}$ the right-hand term is at most $\delta/3$.

Finally, let us prove less than $\delta/3$ the probability that \cardest\ invokes \rs($V$) more than $2\big\lceil\!\sqrt{\kerr n}\,\big\rceil + \kerr$ times.
For convenience let $s=2\big\lceil\!\sqrt{\kerr n}\,\big\rceil$, and let $R(d)$ be the random variable giving the total number of repeats yielded before the $(d+1)$-th distinct sample is obtained.
Note that \cardest\ invokes \rs($V$) more than $s + \kerr$ times if and only if $R(s) < \kerr$.
Let then $\rho_i$ be the indicator random variable of the event that at least one repeat is drawn between the $i$-th and $(i+1)$-th distinct samples.
By construction $R(d) \ge \sum_{i=1}^d \rho_i$ and thus $\prob[R(s) < \kerr] \le \prob[\sum_{i=1}^{s}\rho_i  < \kerr]$.
But $\E[\rho_i] =\frac{i}{n}$, and therefore $\E[\sum_{i=1}^{s}\rho_i] = \frac{s(s+1)}{2} > 2\kerr$.
Therefore $R(s) < \kerr$ implies $\sum_{i=1}^{s}\rho_i < \frac{1}{2}\E[\sum_{i=1}^{s}\rho_i ]$.
Noting that all $\rho_i$ are independent, we can then invoke the following standard concentration bound:
\begin{theorem}[\cite{Dubhashi&2009}, Theorem 1.1]
\label{lem:chernoff}
Let $Y=\sum_{i=1}^n Y_i$ where the $Y_i$ are independently distributed in $[0,1]$. Then for $0<\epsilon<1$ we have $\prob[Y < (1-\epsilon)\E[Y]] \le \exp(-\frac{\epsilon^2}{2}\E[Y])$.
\end{theorem}
\noindent Applying Theorem~\ref{lem:chernoff} to $Y=\sum_{i=1}^{s}\rho_i$, we obtain:
\begin{equation}
\label{eqn_repeatsbound}
\prob\!\big[R\big(s) < \kerr\big]  \le \exp\!\Big(\!-\frac{0.5^2}{2} 2 \kerr \Big)
\end{equation}
and straightforward manipulations show the right-hand term to be less than $\delta/3$.
Since $n+\kerr$ samples always yield at least $\kerr$ repeats, the probability that \cardest\ invokes \rs($V$)  more than $\min(n,2\lceil\!\sqrt{\kerr n}\rceil)+\kerr$ times is less than $\delta/3$.

A simple union bound completes the proof.
\end{proof}

\end{document}